\documentclass[prd,nofootinbib,preprint,superscriptaddress,twocolumn,10pt]{revtex4-1}
\pdfoutput=1
\usepackage{amsmath,amssymb}
\usepackage{epsfig}
\usepackage{graphicx}
\usepackage[usenames,dvipsnames]{color}
\usepackage{subfigure}
\usepackage{slashed}
\usepackage{comment}
\usepackage[normalem]{ulem}
\usepackage[colorlinks,citecolor=blue]{hyperref}
\usepackage{color}

\usepackage{multirow}

\usepackage{xcolor}
\usepackage{soul}

\begin{document}
\title{Probing High Scale Dirac Leptogenesis via Gravitational Waves from Domain Walls}
\author{Basabendu Barman}
\email{basabendu88barman@gmail.com}
\affiliation{Centro de Investigaciones, Universidad Antonio Nariño\\
Carrera 3 este \# 47A-15, Bogotá, Colombia}

\author{Debasish Borah}
\email{dborah@iitg.ac.in}
\affiliation{Department of Physics, Indian Institute of Technology Guwahati, Assam 781039, India}

\author{Arnab Dasgupta}
\email{arnabdasgupta@pitt.edu}
\affiliation{Pittsburgh Particle Physics, Astrophysics, and Cosmology Center, Department of Physics and Astronomy, University of Pittsburgh, Pittsburgh, PA 15206, USA}

\author{Anish Ghoshal}
\email{anish.ghoshal@fuw.edu.pl}
\affiliation{Institute of Theoretical Physics, Faculty of Physics, University of Warsaw, ul. Pasteura 5, 02-093 Warsaw, Poland}
\begin{abstract}
We propose a novel way of probing high scale Dirac leptogenesis, a viable alternative to canonical leptogenesis scenario where the total lepton number is conserved, keeping light standard model (SM) neutrinos purely Dirac. The simplest possible seesaw mechanism for generating light Dirac neutrinos involve heavy singlet Dirac fermions and a singlet scalar. In addition to unbroken global lepton number, a discrete $Z_2$ symmetry is imposed to forbid direct coupling between right and left chiral parts of light Dirac neutrino. Generating light Dirac neutrino mass requires the singlet scalar to acquire a vacuum expectation value (VEV) that also breaks the $Z_2$ symmetry, leading to formation of domain walls in the early universe. These walls, if made unstable by introducing a soft $Z_2$ breaking term, generate gravitational waves (GW) with a spectrum characterized by the wall tension or the singlet VEV, and the soft symmetry breaking scale. The scale of leptogenesis depends upon the $Z_2$-breaking singlet VEV which is also responsible for the tension of the domain wall, affecting the amplitude of GW produced from the collapsing walls. We find that most of the near future GW observatories will be able to probe Dirac leptogenesis scale all the way upto $10^{11}$ GeV.
\end{abstract}
\begin{flushright}
PI-UAN-2022-716FT
\end{flushright} 
\maketitle
\noindent
\textbf{\textit{Introduction:}} The observed universe is asymmetric in its baryon content with excess of baryons over antibaryons, quoted in terms of baryon to photon ratio as~\cite{Aghanim:2018eyx} 
\begin{equation}
\eta_B = \frac{n_{B}-n_{\bar{B}}}{n_{\gamma}} = 6.1 \times 10^{-10}\,,
\label{etaBobs}
\end{equation}
based on cosmic microwave background (CMB) measurements. This observed asymmetry which is also consistent with the big bang nucleosynthesis (BBN) predictions has been a longstanding puzzle, as the standard model (SM) can not provide a viable explanation. While the SM fails to fulfill the Sakharov's conditions~\cite{Sakharov:1967dj} necessary for generating baryon asymmetry dynamically, several beyond standard model (BSM) frameworks consider out-of-equilibrium decay of heavy particles leading to baryogenesis~\cite{Weinberg:1979bt, Kolb:1979qa}. One appealing alternative, known as leptogenesis~\cite{Fukugita:1986hr}, is to generate such an asymmetry in the lepton sector first which later gets converted into baryon asymmetry through $(B+L)$-violating electroweak (EW) sphaleron transitions~\cite{Kuzmin:1985mm}. While most of the leptogenesis scenarios consider Majorana nature of light neutrinos, one equally appealing alternative is to consider Dirac nature of light neutrinos. As proposed in~\cite{Dick:1999je, Murayama:2002je}, one can have successful leptogenesis even with light Dirac neutrino scenarios where total lepton number or $B-L$ is conserved just like in the SM. Popularly known as the Dirac leptogenesis scenario, it involves the creation of an equal and opposite amount of lepton asymmetry in left handed and right handed neutrino sectors followed by the conversion of left sector asymmetry into baryon asymmetry via electroweak sphalerons. The lepton asymmetries in the left and right handed sectors are prevented from equilibration due to the tiny effective Dirac Yukawa couplings. Various possible implementation of this idea can be found in, for example, Refs.~\cite{Boz:2004ga, Thomas:2005rs, Thomas:2006gr, Cerdeno:2006ha, Gu:2006dc, Gu:2007mc, Chun:2008pg, Bechinger:2009qk, Chen:2011sb, Choi:2012ba, Borah:2016zbd, Gu:2016hxh, Narendra:2017uxl}. In a few related works~\cite{Heeck:2013vha, Gu:2019yvw, Mahanta:2021plx}, violation of $B-L$ symmetry was accommodated in a way to preserve the Dirac nature of light neutrinos, while generating lepton asymmetry simultaneously. 

Irrespective of Dirac or Majorana nature of neutrinos, the leptogenesis, in general, is a high scale phenomena with very limited scope of observational signatures. Recently, one interesting possibility of probing high scale leptogenesis via stochastic gravitational wave (GW) observation was pointed out in~\cite{Dror:2019syi} followed by a few related works in Refs.~\cite{Blasi:2020wpy, Fornal:2020esl, Samanta:2020cdk}. These works assume the presence of an Abelian gauge symmetry like $U(1)_{B-L}$ with a Type-I seesaw framework, such that the breaking of this symmetry can lead to the formation of cosmic strings that emit GW, and also dictate the mass of heavy Majorana right handed neutrinos, i.e., the scale of leptogenesis. For high scale $U(1)_{B-L}$ breaking where vanilla leptogenesis is valid, the GW spectrum from cosmic strings lies within current and next generation experimental sensitivities. While the central key assumption behind these analyses is the presence of such additional gauge symmetry, which is not necessary for the validity of Type-I seesaw and high scale leptogenesis, we consider, in this paper, a Dirac leptogenesis scenario where a discrete symmetry like $Z_2$ is a must to validate the seesaw origin of light Dirac neutrino mass. In order to generate the light Dirac neutrino mass via Dirac seesaw, this $Z_2$ symmetry gets spontaneously broken by a singlet scalar field acquiring non-zero vacuum expectation value (VEV), which in turn, also leads to the formation of topological defects known as domain walls (DW) in the early universe. These DW can become metastable in the presence of explicit $Z_2$ breaking term that induces a pressure difference (also known as {\it{bias term}}) across the walls. Such metastable DW can emit stochastic GW whose spectrum is dependent on the singlet VEV, as well as the bias term. Since the singlet VEV also controls the Yukawa couplings and heavy neutral lepton mass in Dirac seesaw, we get interesting correlations between the scale of Dirac leptogenesis and GW spectrum that can be probed at several proposed GW detectors. Additionally, since GW spectrum from DW network is distinctly different from the one generated by cosmic strings, it also provides an interesting way to distinguish Dirac leptogenesis from Majorana leptogenesis studied in earlier work \cite{Dror:2019syi} in the context of GW probe. This is complementary to the neutrinoless double beta decay (NDBD) \cite{Dolinski:2019nrj} probe of Majorana neutrinos. Therefore, our scenario can not only be probed at future GW experiments but also remains falsifiable by future observations of NDBD. While non-observation of NDBD does not necessarily confirm Dirac nature of light neutrinos, future observation of neutrinoless quadrupole beta decay \cite{Heeck:2013rpa} can confirm it, providing another way of testing our setup.


\medskip
\noindent
\textbf{\textit{Dirac Leptogenesis:}} We consider the most minimal seesaw realisation of light Dirac neutrinos. The SM is extended by three copies of heavy Dirac fermions $N_{L,R}$, right handed counter part of active neutrinos $\nu_R$ and a singlet scalar $\eta$. In order to ensure pure Dirac nature of light neutrinos, a global lepton number symmetry is assumed, which can also be remnant after some gauge symmetry breaking, not considered in this minimal setup. In order to prevent direct coupling of the SM lepton doublet $L$ with $\nu_R$ via SM Higgs $H$, a discrete $Z_2$ symmetry is imposed under which $\eta, \nu_R$ are odd while all other fields are even. The relevant Yukawa Lagrangian then reads
\begin{align}
    -\mathcal{L}_Y \supset Y_L\, \overline{L}\,\tilde{H}\,N_R + M_N\, \overline{N}\,N + Y_R\,\overline{N_L}\,\eta \,\nu_R + {\rm h.c.}
\end{align}
After the neutral components of $H$ and $\eta$ acquire VEV $v,\, u$ respectively, light Dirac neutrino mass arises from the Type-I seesaw equivalent for Dirac neutrino as
\begin{equation}
    M_{\nu} = \frac{1}{\sqrt{2}}Y_L\,M^{-1}_N\,Y_R\,v\,u\,.
\end{equation}
While the net lepton asymmetry is zero due to unbroken lepton number symmetry, one can still create equal and opposite lepton asymmetries in left and right sectors due to CP violating out-of-equilibrium decays $N \rightarrow L\,H$ and $N \rightarrow \nu_R\,\eta$ respectively. 
The masses of the Dirac fermions $N$'s are taken to be of the same order with the mass matrix in diagonal form as $M_N = {\rm Diag} (M_1,2M_1,3M_1)$ throughout our analysis. The $CP$ asymmetry parameter is given as \cite{Cerdeno:2006ha}
\begin{align}
    \epsilon \simeq -\frac{1}{8\pi}\frac{M_1}{uv}\frac{{\rm  Im}[(Y_R m^{\dagger}_{\nu}Y_L)_{11}]}{(Y_RY^{\dagger}_R) + (Y_LY^{\dagger}_L)},
    \label{eq:cp}
\end{align}
where $v = 246$ GeV. Now the effective neutrino mass ($\widetilde{m}$) is defined as 
\begin{align}
    \widetilde{m} &= [(Y_RY^{\dagger}_R) + (Y_LY^{\dagger}_L)]\frac{uv}{M_1}
\end{align}
and without loss of generality we have assumed $Y_L\sim Y_R = y$. Now, plugging in the above equation back into Eq. \eqref{eq:cp} we parametrise the CP asymmetry as
\begin{align}
    \epsilon &= \frac{1}{8\pi} y^2 \sin(2\phi)\frac{m_\nu}{\widetilde{m}} \simeq \frac{1}{8\pi} y^2 \sin(2\phi).
\end{align}

During the sphaleron transitions, the asymmetry in the left sector can get converted into a net baryon asymmetry. This, however, depends on the condition that the asymmetries in the left and right sectors do not equilibrate with each other. Such processes leading to equilibration of left and right sector asymmetries at high temperature can be approximated to be
\begin{align}
\Gamma_{L-R} &\sim \frac{|Y_L|^2|Y_R|^2}{M^2_1}T^3\,,
\end{align}
which in turn should be less than the Hubble expansion rate at the radiation dominated era given as
\begin{align}
\mathcal{H}(T) &= \sqrt[]{\frac{8\pi^2g_*}{90}}\frac{T^2}{M_{\rm Pl}}
\end{align}
The strongest bounds comes from the high temperature when the asymmetry is produced i.e $z =M_1/T \simeq 1$
\begin{align}
\frac{|Y_L|^2|Y_R|^2}{M_1} &\leq \frac{1}{M_{\rm Pl}}\sqrt[]{\frac{8\pi^2 g_*}{90}}
\end{align}
We follow the recipe outlined in~\cite{Cerdeno:2006ha, Borah:2016zbd} and numerically solve the relevant Boltzmann equations (BEQ) for calculating the final asymmetry, considering the heavy neutral fermions to be hierarchical in masses.

\begin{figure}
\includegraphics[scale=0.5]{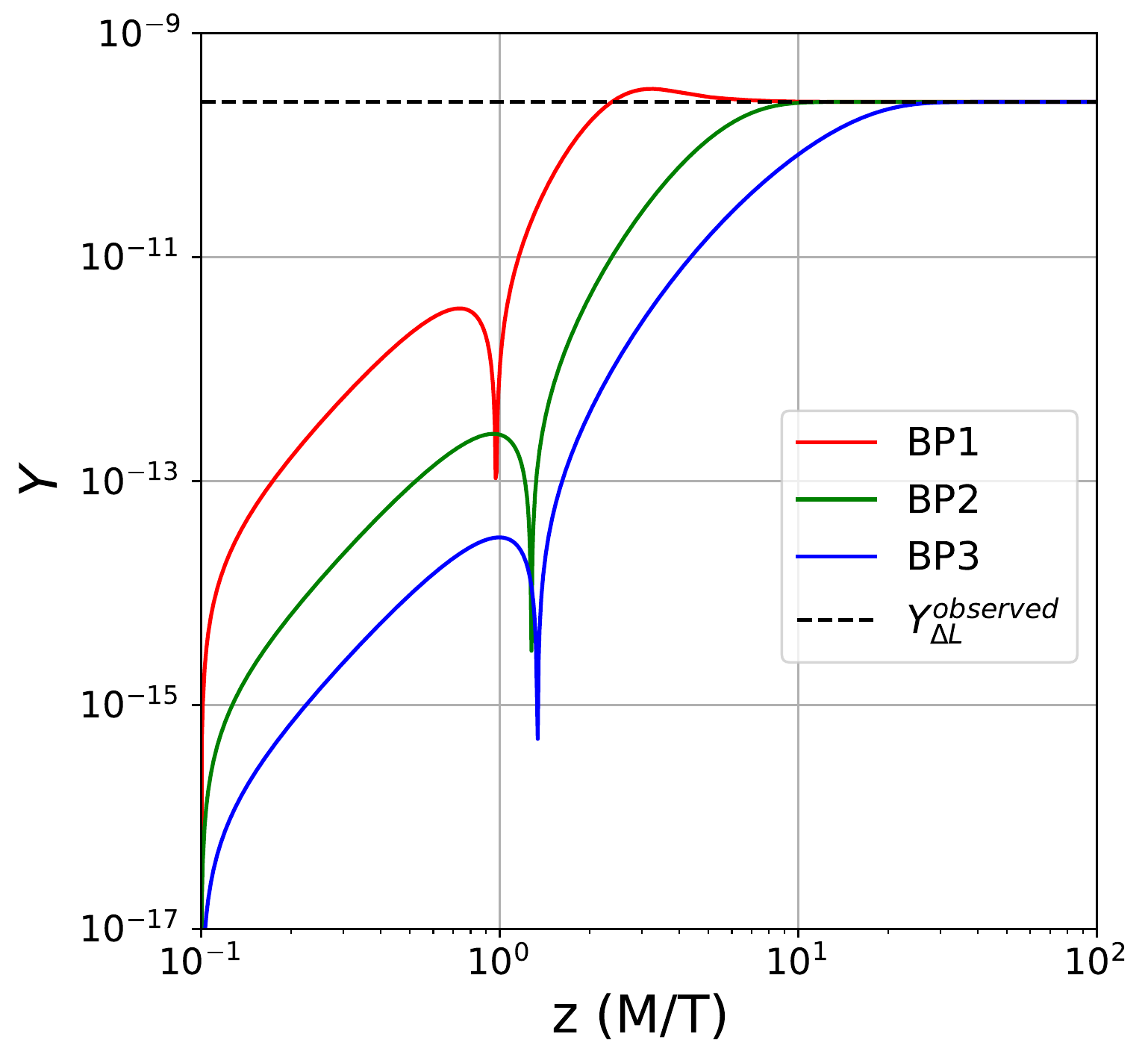}
\caption{Evolution of lepton asymmetry for different benchmark parameters shown in table \ref{tab:Lepto}. The dashed horizontal line corresponds to the required lepton asymmetry which can be converted into the observed baryon asymmetry by sphalerons.}
\label{fig:lepto} 
\end{figure} 

\begin{figure}
\includegraphics[scale=0.58]{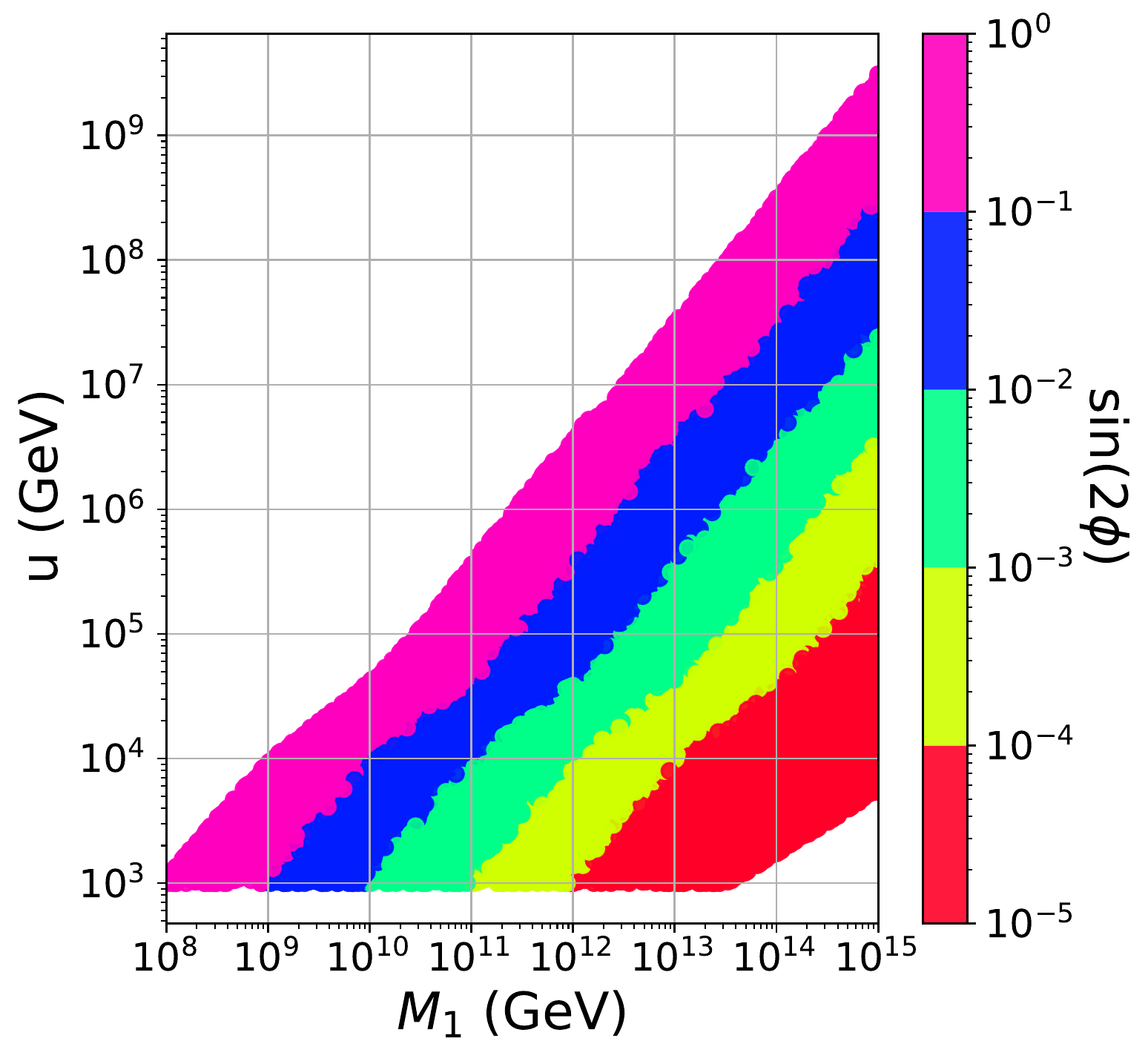}
\caption{Parameter space consistent with correct baryon asymmetry.}
\label{fig:lepto2} 
\end{figure} 

\begin{table}
    \centering
    \begin{tabular}{|c|c|c|c|c|}
    \hline
     & $M_{1}$ (GeV)  & $u$ (GeV) & $Y_L=Y_R$ & $\sin (2\phi)$\\
    \hline  BP1 & $10^{12}$  & $10^4$ & $4.51 \times 10^{-3}$ & -$1.12\times 10^{-3}$ \\
    BP2 & $10^{12}$   & $ 10^5$ & $1.43\times 10^{-3}$ & -$2.84\times 10^{-2}$\\
    BP3 & $10^{12}$   & $10^6$ & $4.51\times 10^{-4}$ & -$0.266$\\
    \hline
    \end{tabular}
    \caption{Details of the benchmark parameters used to depict the evolution of lepton asymmetry.}
  \label{tab:Lepto}
\end{table}

In Fig.~\ref{fig:lepto}, we show the evolution of comoving density of lepton number in either of the sectors as a function of $z=M_1/T$ for three different benchmark choices of parameters given in Tab.~\ref{tab:Lepto}. While the scale of leptogenesis is kept the same, variation is shown for different singlet VEV (and hence the Yukawa couplings) as required from satisfying the light neutrino mass constraints. For simplicity, the left and right sector Yukawa couplings are considered to be the same. Denoting the effective CP phase entering the CP asymmetry formula as $\sin{2\phi}$, we perform a numerical scan for singlet VEV and the scale of leptogenesis $M_1$ which is consistent with the required lepton asymmetry while keeping $Y_L = Y_R$ which get restricted from light neutrino data. The viable parameter space is shown in Fig.~\ref{fig:lepto2}. As expected, a larger value of singlet VEV for fixed CP phase corresponds to larger $M_1$ in order to satisfy light Dirac neutrino mass criteria. Additionally, for a fixed value of singlet VEV, a lower value of $M_1$ requires larger $\sin{2\phi}$ in order to generate sufficient CP asymmetry. 

\begin{figure}
\includegraphics[scale=0.4]{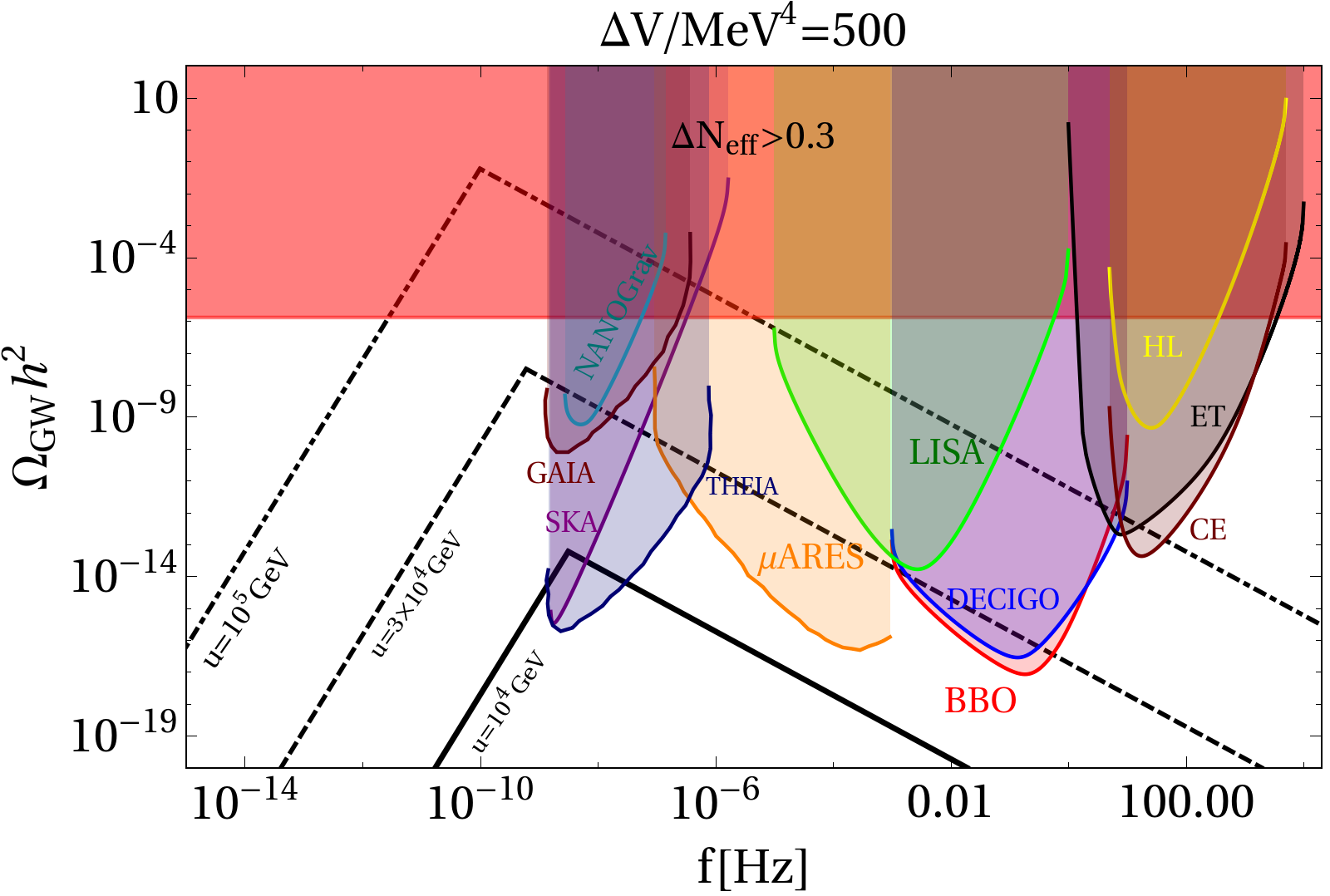}
\caption{Gravitational wave spectrum from domain walls, where different straight black lines correspond to different choices of $u$ that is consistent with baryon asymmetry, while different coloured curves show the sensitivities from GW search experiments like LISA, BBO, DECIGO, HL (aLIGO), ET, CE, NANOGrav, SKA, GAIA, THEIA and $\mu$ARES. The shaded region parallel to the horizontal axis is excluded from the fact that DW network becomes long-lived enough to dominate the energy density of the universe before collapsing and emits large amount of radiation violating PLANCK bounds on $\Delta N_\text{eff}$.}
\label{fig:gw} 
\end{figure} 

\begin{figure*}
\centering
\begin{tabular}{lcr}
\includegraphics[scale=0.35]{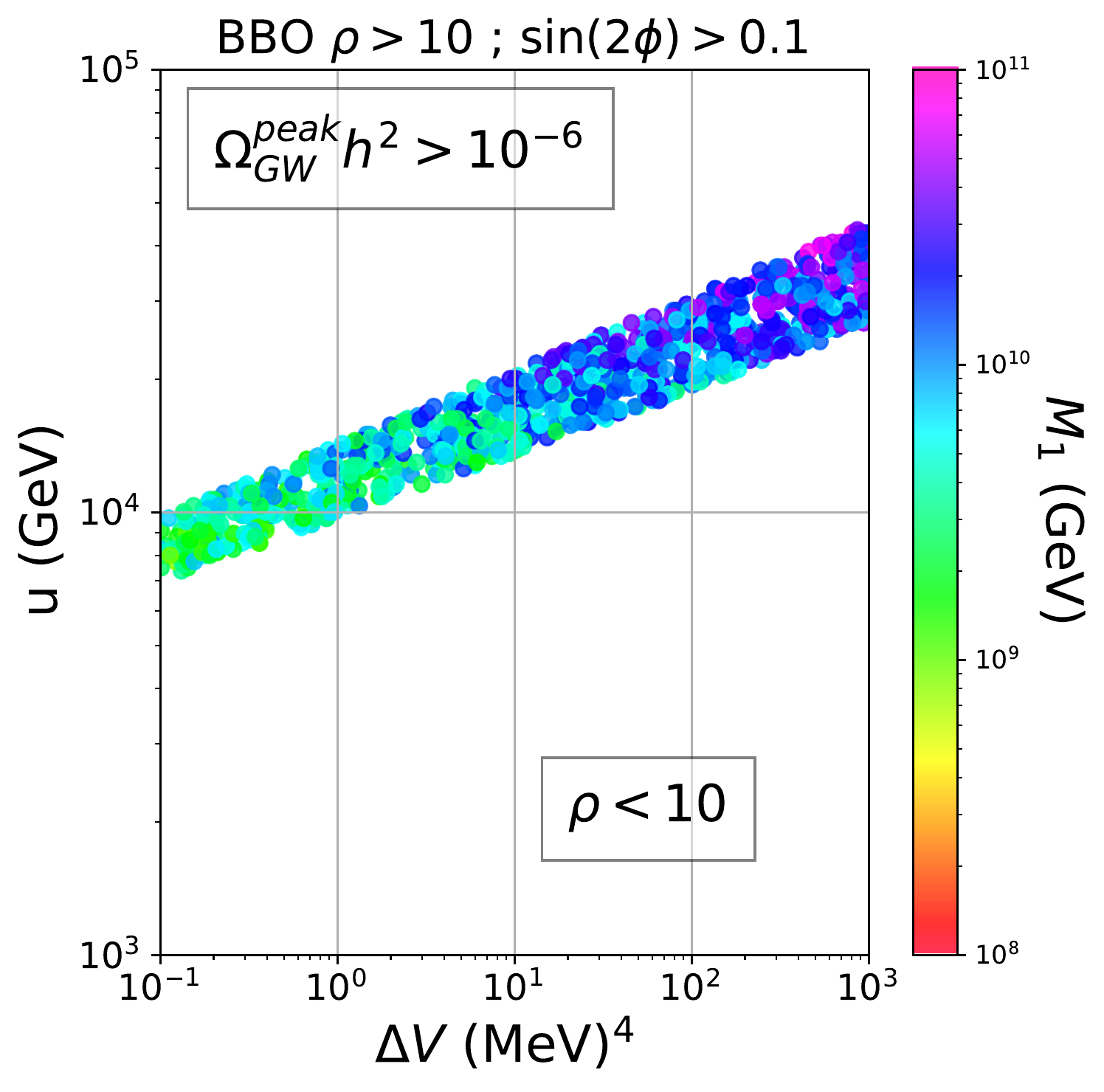}     & 
 \includegraphics[scale=0.35]{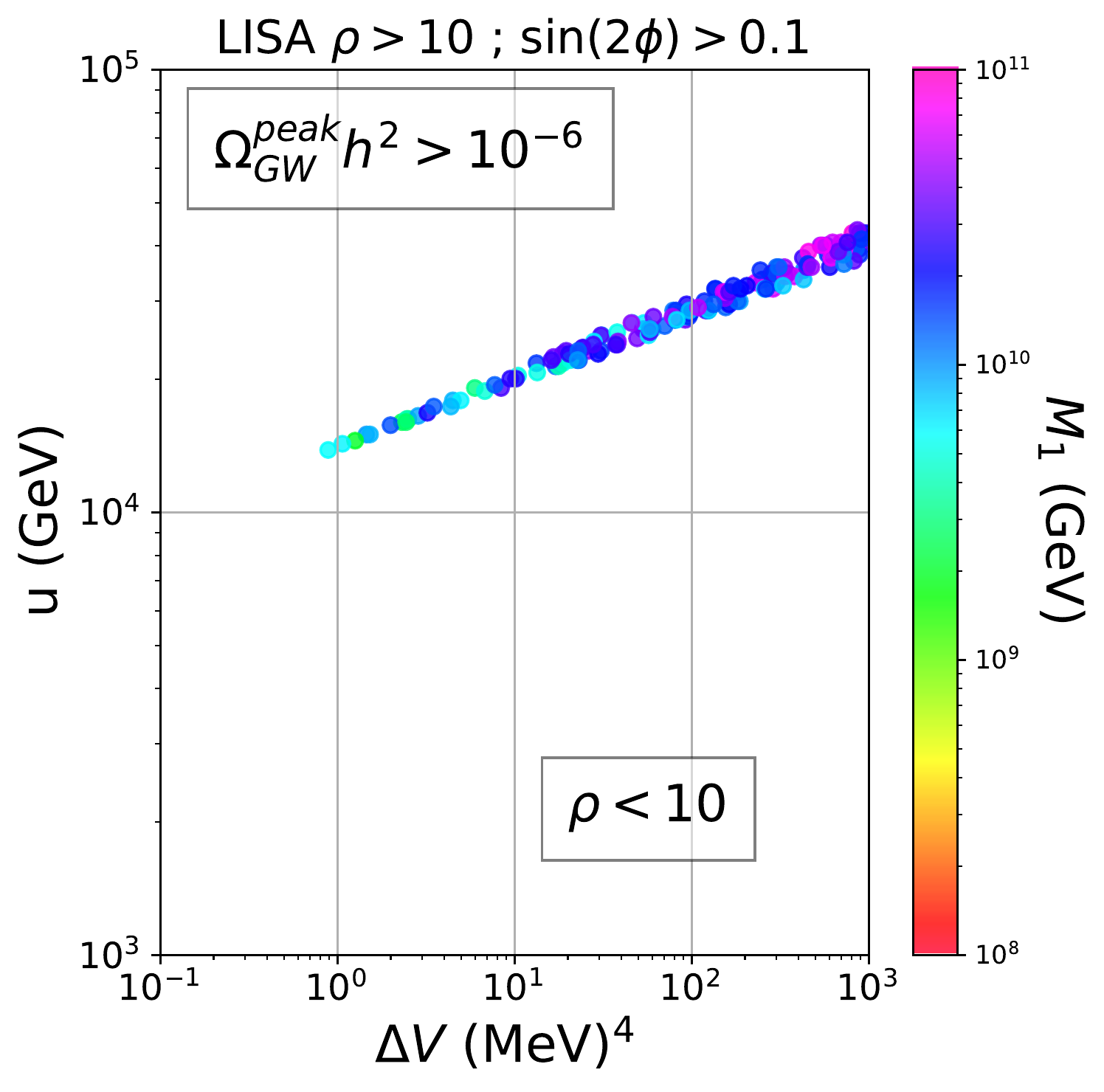}    & 
 \includegraphics[scale=0.35]{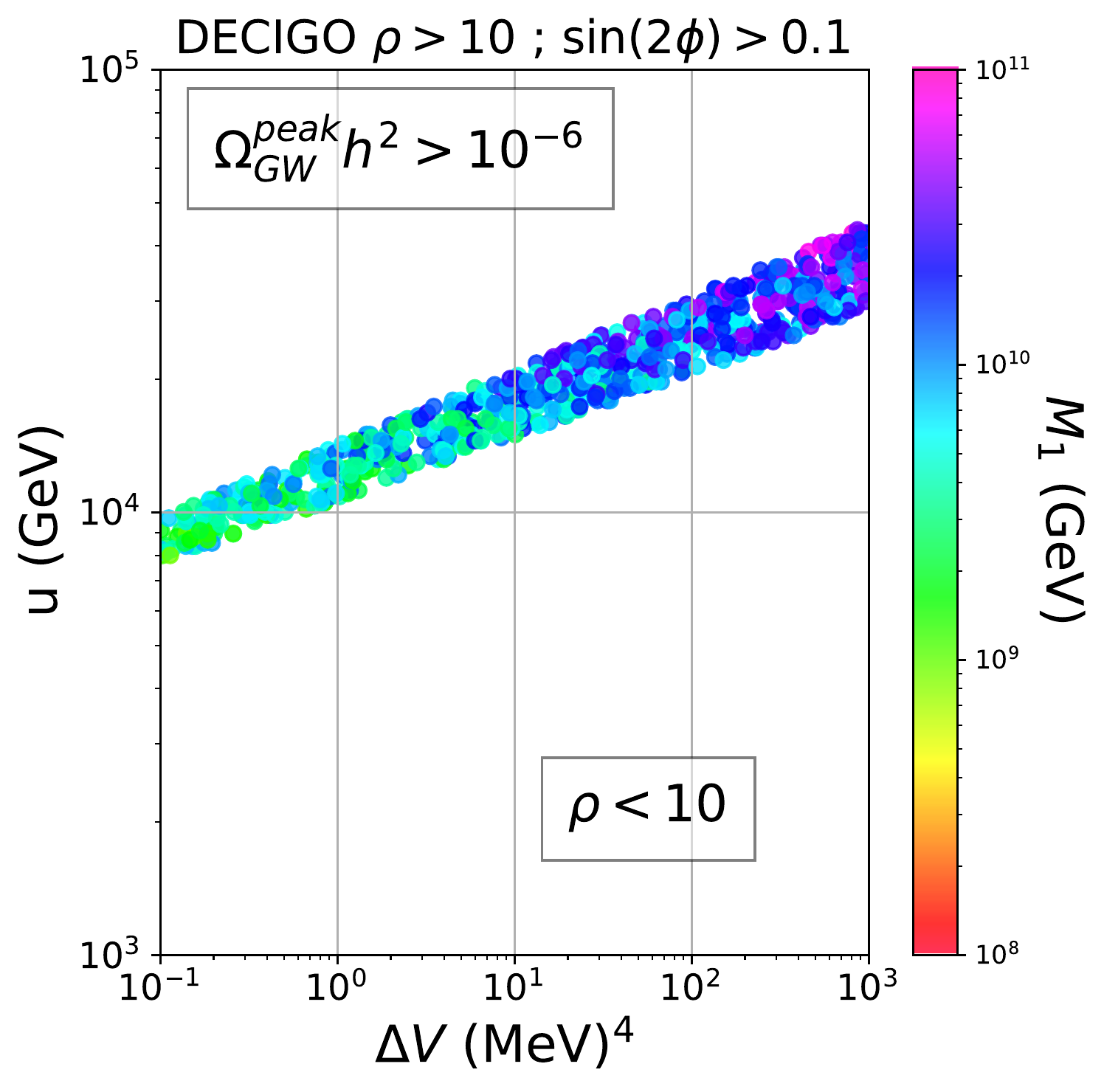} \\
 \includegraphics[scale=0.35]{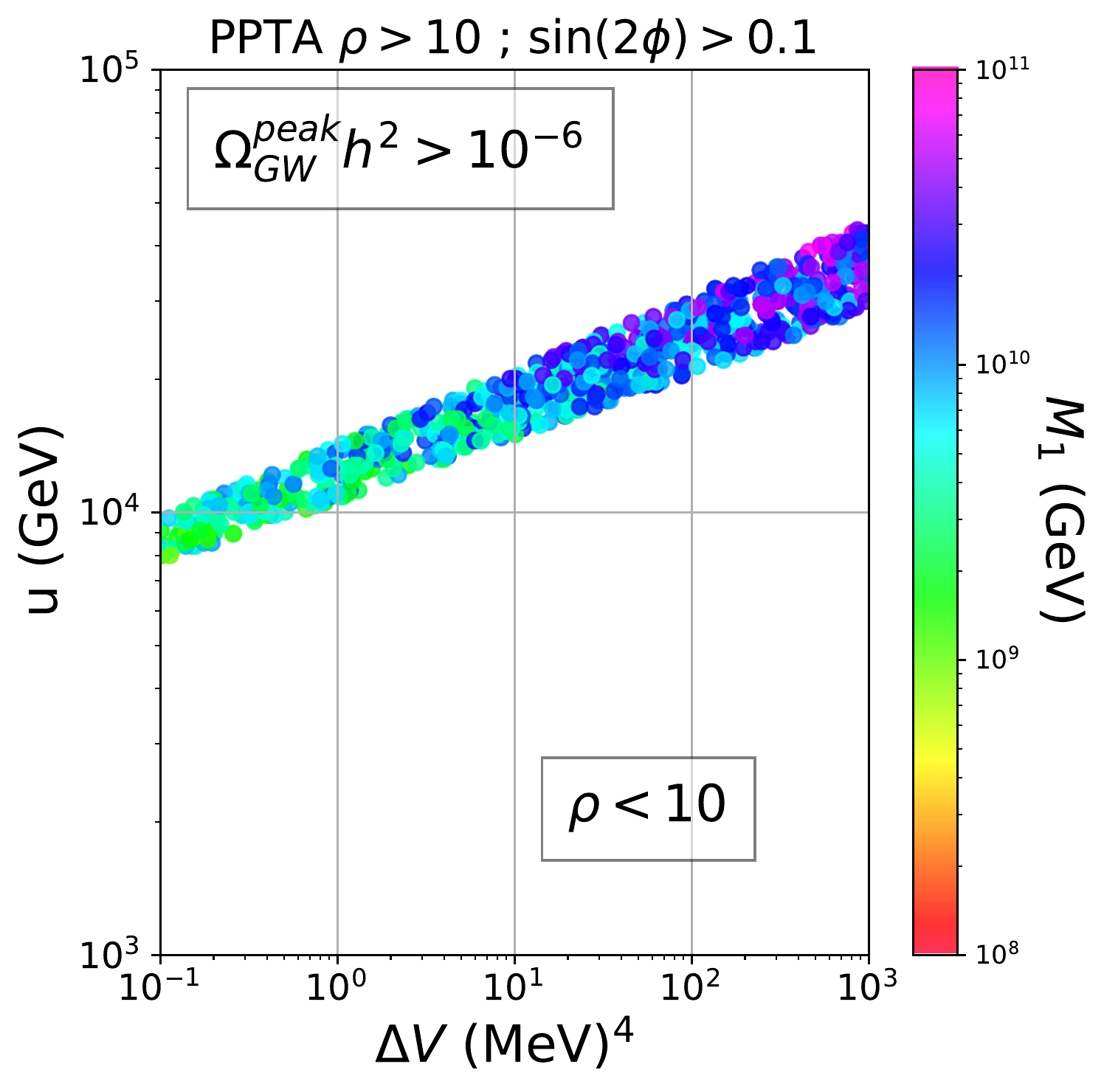}     & 
 \includegraphics[scale=0.35]{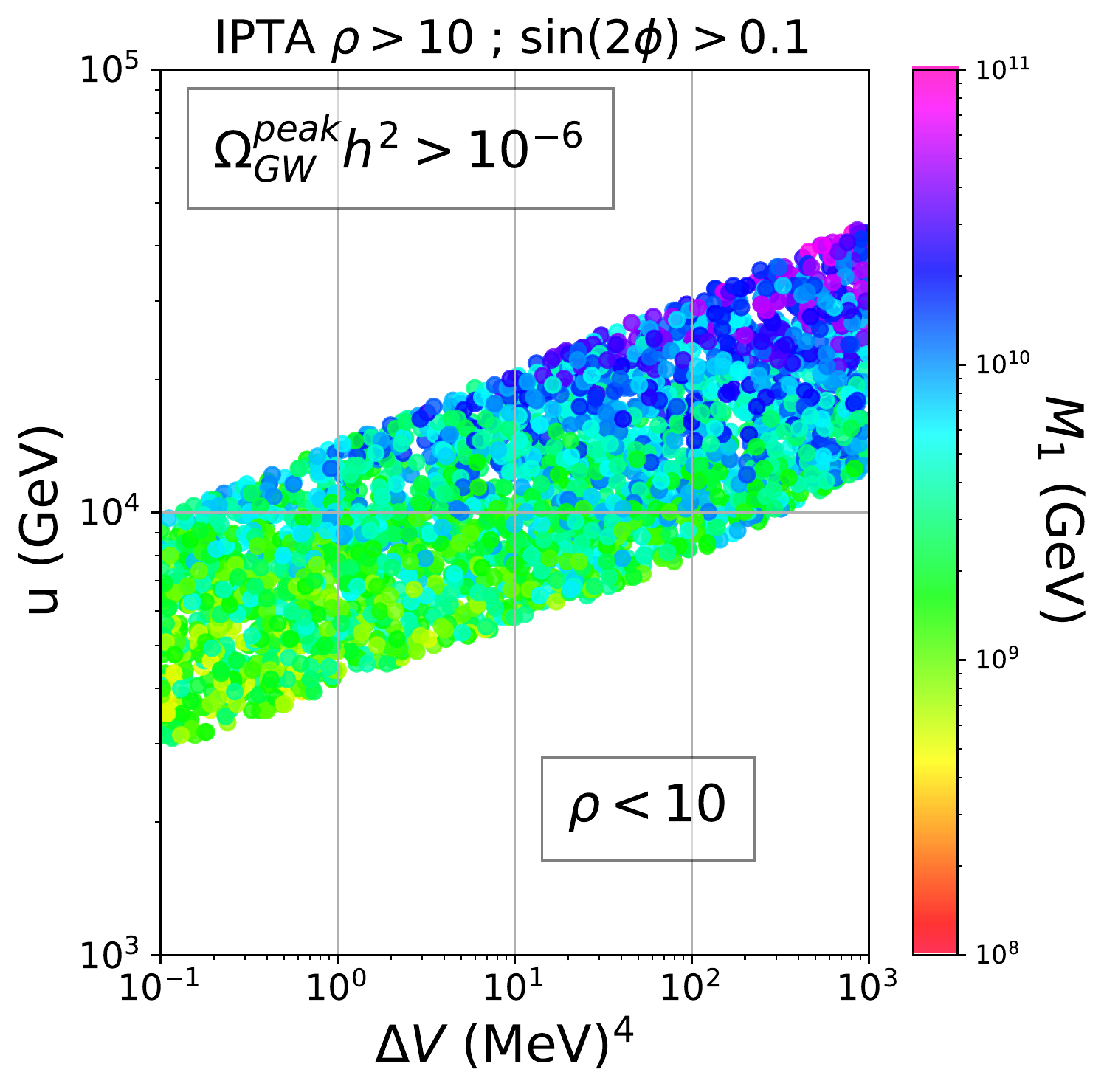}    & 
 \includegraphics[scale=0.35]{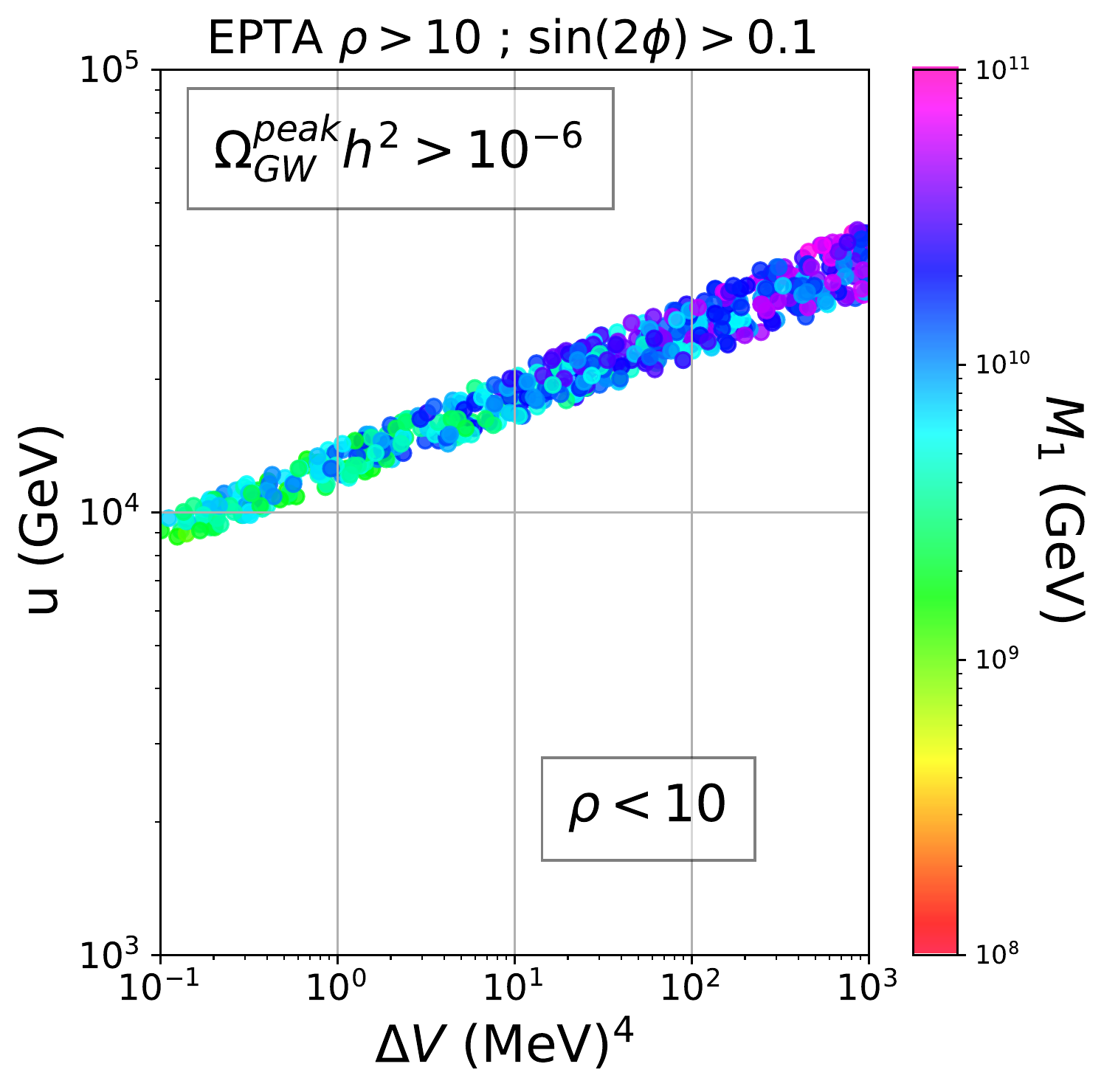} \\
 \includegraphics[scale=0.35]{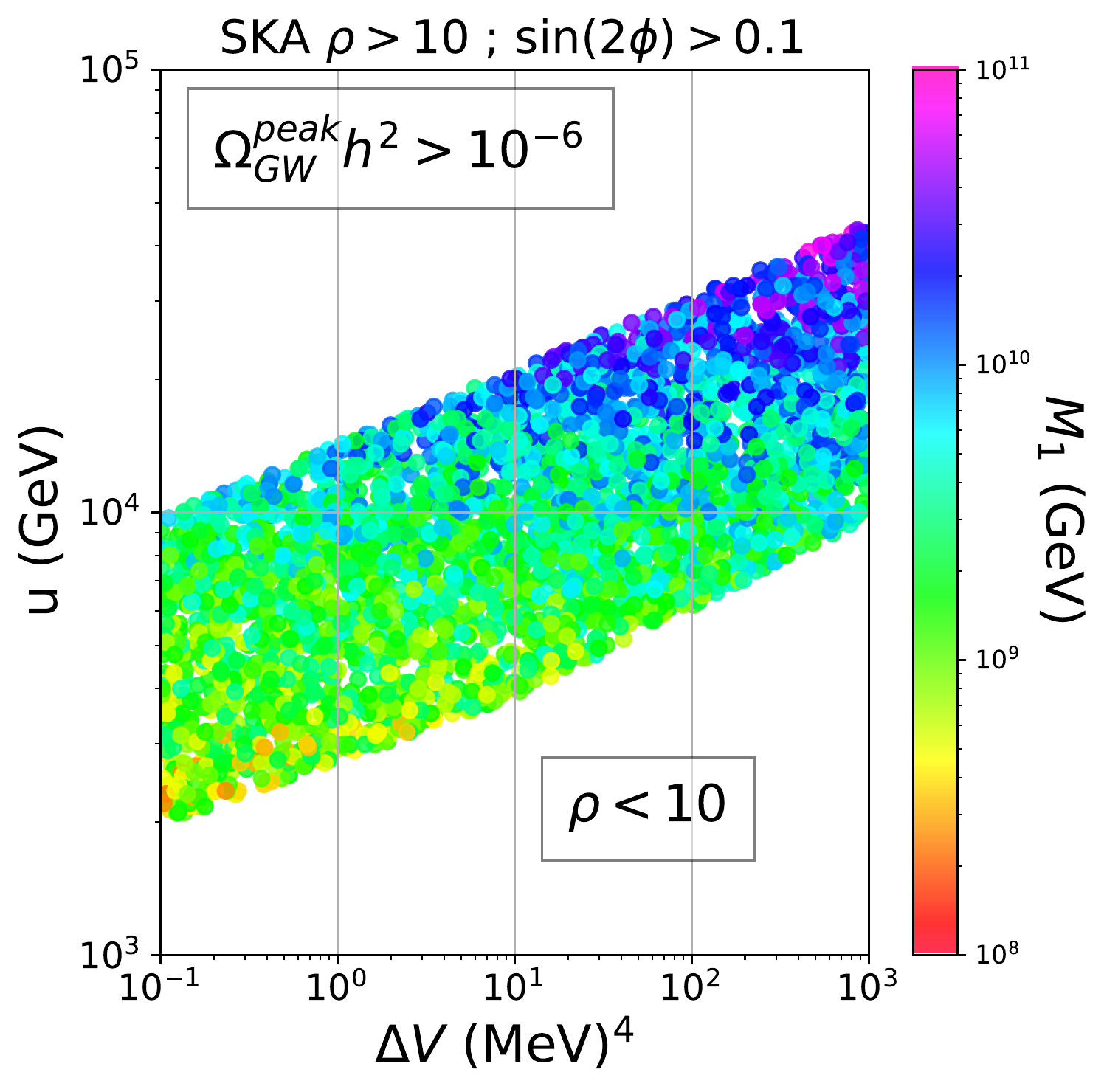} &
 \includegraphics[scale=0.35]{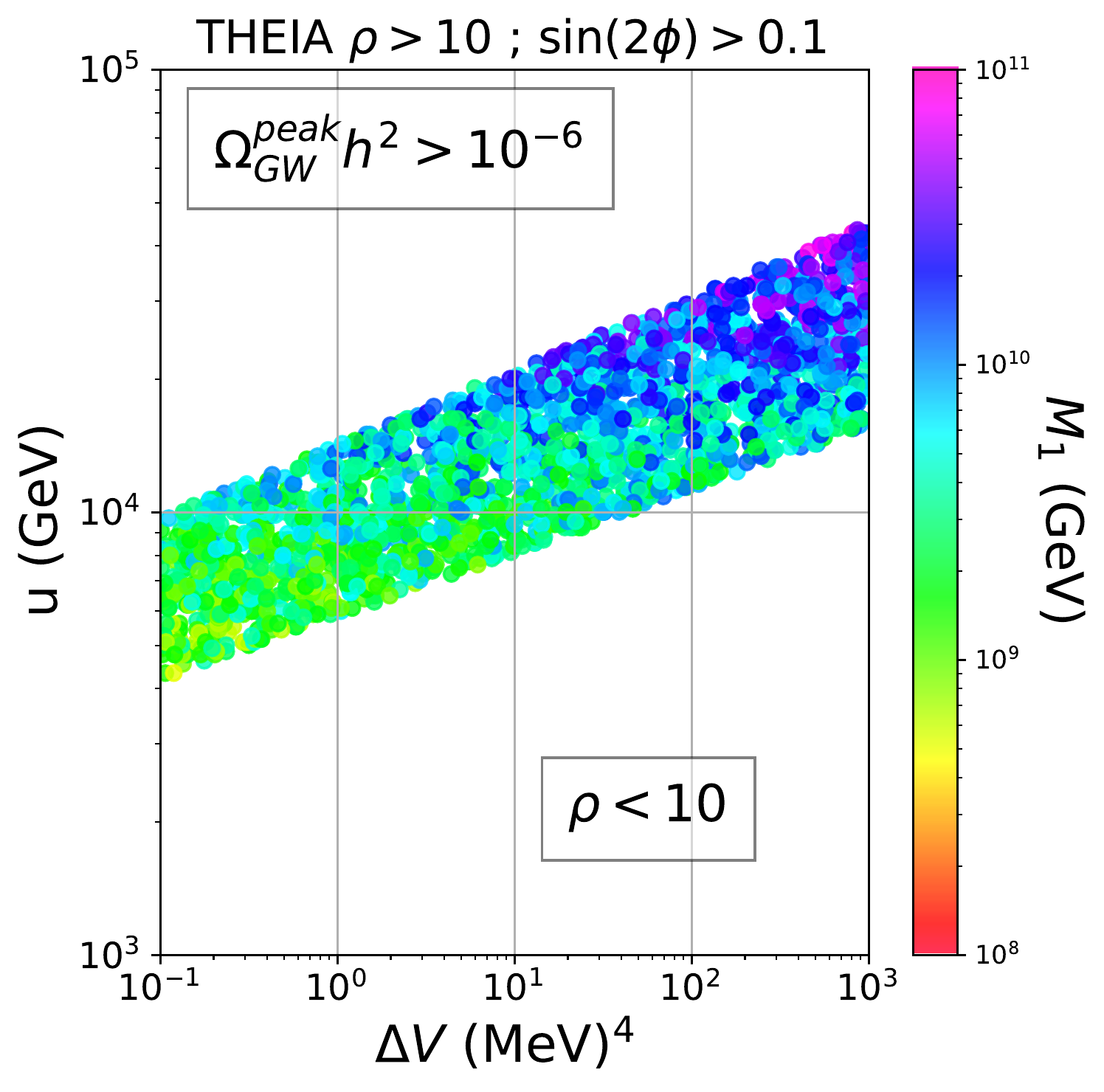} &
  \includegraphics[scale=0.35]{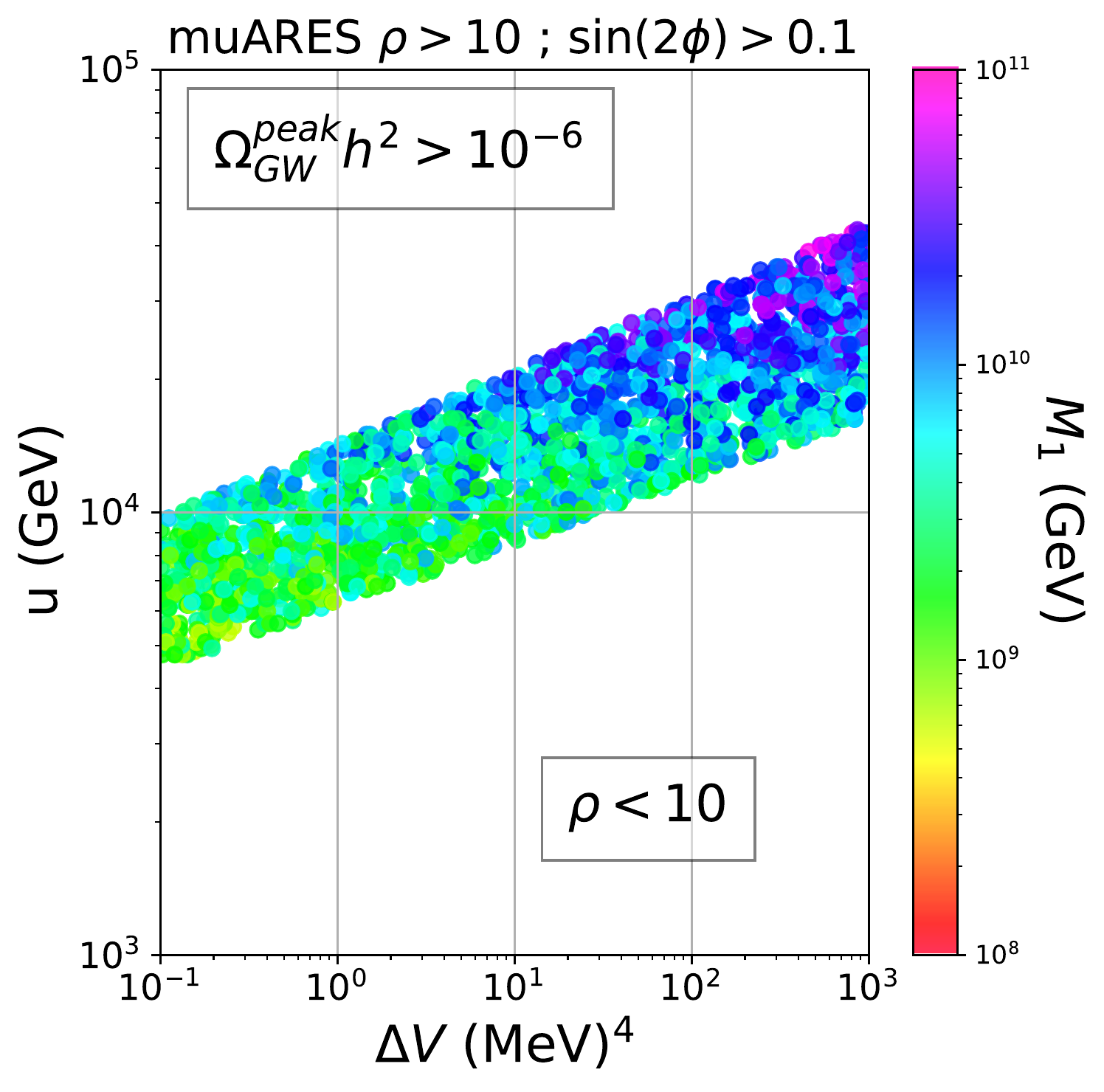}
\end{tabular}
  \includegraphics[scale=0.35]{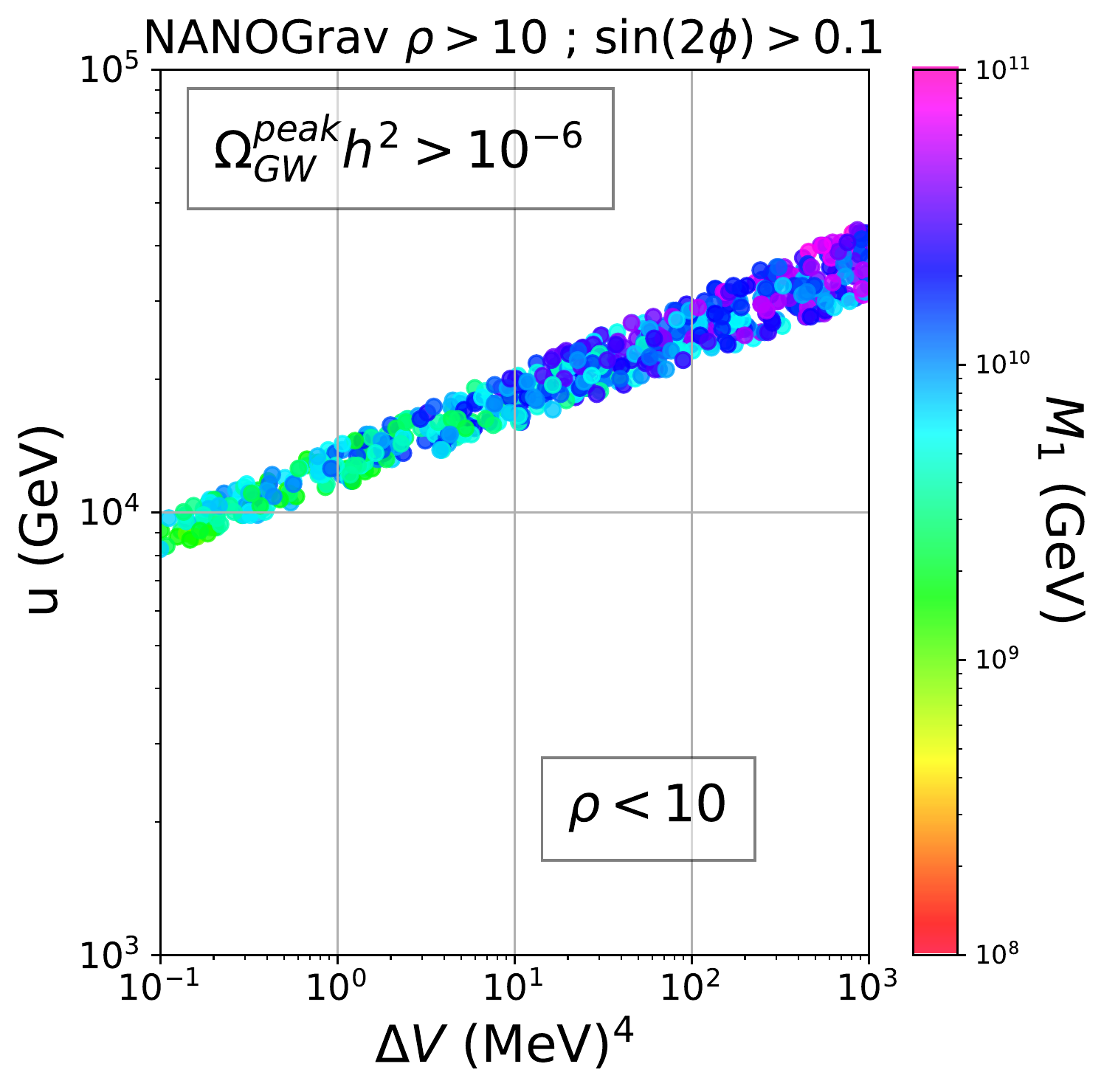}
\caption{Parameter space in singlet VEV $u$ versus bias term $\Delta V$ with leptogenesis scale in colour code. The CP phase parameter $ \sin{2\phi}>0.1 $ and the signal to noise ratio for respective experiments is taken to be more than 10. The region above the coloured patch is ruled out by cosmological limits from BBN as well as CMB observations while the region below the coloured patch corresponds to SNR $<$ 10.}
\label{fig:cntr-SNR} 
\end{figure*} 

\medskip
\noindent
\textbf{\textit{Domain walls and gravitational waves:}} Spontaneous breaking of discrete symmetries in the early universe can lead to the formation of topological defects like domain walls~\cite{Zeldovich:1974uw, Kibble:1976sj, Vilenkin:1981zs}. The energy density of domain walls, which form after spontaneous breaking of discrete symmetries, redshifts slower compared to matter or radiation and can start dominating the energy density of the universe at some epoch. In order to prevent DW to dominate the energy density of the universe, the walls need to be unstable or diluted or if the probability distribution for initial field fluctuations is asymmetric \cite{Coulson:1995nv, Krajewski:2021jje}. In our setup with a $Z_2$-odd scalar singlet $\eta$, having the potential
\begin{align}
	V(\eta) = \frac{\lambda_\eta}{4}(\eta^2-u^2)^2\,,
	\label{V}
\end{align}
it is possible to find a static solution of the equation of motion after imposing a boundary condition such that the two vacua are realized at $x \to \pm \infty$,
\begin{align}
	\eta({\bf x}) = u \tanh\left( \sqrt{\frac{\lambda_\eta}{2}} u x \right).
\end{align}
The above solution represents a domain wall extended along the $x = 0$ plane. The DW width $\delta$ is approximately the inverse of the mass of $\eta$ at the potential minimum: $\delta \sim m_\eta^{-1} = (\sqrt{2\lambda_\eta} u)^{-1}$. Another key parameter for DW, known as the DW tension is given by
\begin{align}
	\sigma = \int_{-\infty}^{\infty} dx \,\rho_\eta = \frac{2\sqrt 2}{3}\sqrt{\lambda_\eta} u^3 = \frac{2}{3}m_\eta u^2\,,
\end{align}
where $\rho_\eta = \frac{1}{2} |\nabla \eta|^2 + V(\eta)$ is the (static) energy density of $\eta$. For $m_\eta \sim u$, the tension of the wall can be approximated as $\sigma \sim u^3$.

The walls can be made unstable simply by introducing a pressure difference across the walls, a manifestation of a small explicit symmetry breaking term~\cite{Zeldovich:1974uw, Vilenkin:1981zs, Sikivie:1982qv, Gelmini:1988sf}. Such a pressure difference or bias term $\Delta V$ should be large enough such that the DW do not start dominating the universe and disappear at least before the epoch of BBN, in order not to disturb the success of standard cosmology. On the other hand, the bias term $\Delta V$ can not be arbitrarily large due to the requirement of percolation of both the vacua (separated by DW) whose relative population can be estimated as $p_+/p_- \simeq e^{-4\Delta V/(\lambda_\eta u^4)}$ \cite{Gelmini:1988sf}. Such unstable DW can emit gravitational waves, the details of which has been studied in several works~\cite{Kadota:2015dza, Hiramatsu:2013qaa, Krajewski:2016vbr, Nakayama:2016gxi, Dunsky:2020dhn, Babichev:2021uvl, Ferreira:2022zzo, Deng:2020dnf, Gelmini:2020bqg, Saikawa:2017hiv}. The amplitude of such GW at peak frequency $f_{\rm peak}$ can be estimated as~\cite{Kadota:2015dza, Hiramatsu:2013qaa}
\begin{align}
    \Omega_{\rm GW}h^2 (t_0) \rvert_{\rm peak} & \simeq 5.2 \times 10^{-20} \tilde{\epsilon}_{\rm gw} A^4 \left ( \frac{10.75}{g_*} \right)^{1/3} \nonumber \\
    & \times \left ( \frac{\sigma}{1 \, {\rm TeV}^3 } \right)^4  \left ( \frac{1 \, {\rm MeV}^4}{\Delta V} \right)^2\,,
\end{align}
with $t_0$ being the present time. Away from the peak, the amplitude varies as 
\begin{align}
	\Omega_{\rm GW} \simeq \Omega_{\rm GW}\rvert_{\rm peak} \times 
	\begin{cases}
		\displaystyle{\left( \frac{f_{\rm peak}}{f} \right)} & {\rm for}~~f>f_{\rm peak}\\
		&\\
		\displaystyle{\left( \frac{f}{f_{\rm peak}} \right)^3 }& {\rm for}~~f<f_{\rm peak}
	\end{cases}\,,
\end{align}
where the peak frequency reads
\begin{align}
    f_{\rm peak} (t_0) & \simeq 3.99 \times 10^{-9} \, {\rm Hz} A^{-1/2} \nonumber \\
    & \times \left ( \frac{ 1\, {\rm TeV}^3}{\sigma} \right)^{1/2} \left ( \frac{\Delta V}{1\, {\rm MeV}^4} \right)^{1/2}\,.
\end{align}
In the above expressions, $A$ is the area parameter~\cite{Caprini:2017vnn, Paul:2020wbz} $\simeq 0.8$ for DW arising from $Z_2$ breaking, and $\tilde{\epsilon}_{\rm gw}$ is the efficiency parameter $\simeq$ 0.7~\cite{Hiramatsu:2013qaa}. Since the GW amplitude at peak frequency increases with DW tension or equivalently, the singlet scalar VEV, we need to consider an upper bound such that the resulting GW do not dominate the energy density of the universe. For example, cosmological observations from the PLANCK satellite and the corresponding CMB limits on additional effective relativistic degrees of freedom $\Delta N_{\rm eff}$ can be used to put upper bound $\Omega_{\rm GW} h^2 \lesssim 10^{-6}$~\cite{Boyle:2007zx,Stewart:2007fu,Pagano:2015hma, Lasky:2015lej, Aghanim:2018eyx}. Similar bounds can be applied from the BBN limits on $\Delta N_{\rm eff}$ as well. It should be noted that we are ignoring the friction effects between the walls and the thermal plasma \cite{Nakayama:2016gxi, Galtsov:2017udh} which can be significant if the field constituting the wall has large couplings with the SM bath particles like the Higgs. In the presence of such friction effects, the amplitude of GW emitted by the collapsing walls will be smaller than that without friction discussed here. We neglect such frictional effects assuming the singlet scalar coupling with the SM bath to be tiny \cite{Babichev:2021uvl}.

In Fig.~\ref{fig:gw} we show the GW spectrum arising from DW by choosing some benchmark values of singlet scalar VEV $u$ while keeping the bias term fixed at $\Delta V = 500 \, {\rm MeV}^4$. As expected, with increase in singlet VEV, the DW tension also rises enhancing the GW amplitude. For the chosen benchmark points, only one of the peak frequencies remain within the experimental sensitivities while the region of higher frequencies for all the benchmark points remain within reach of experiments. Very large values of $u$ for the chosen bias term get disfavoured by the upper bound on GW amplitude from cosmology data, shown as the pink shaded region in the uppermost region. The experimental sensitivities of NANOGrav \cite{McLaughlin:2013ira} , SKA \cite{Weltman:2018zrl}, GAIA \cite{Garcia-Bellido:2021zgu}, THEIA \cite{Garcia-Bellido:2021zgu}, $\mu$ARES \cite{Sesana:2019vho}, LISA\,\cite{AmaroSeoane2012LaserIS}, DECIGO \cite{Kawamura:2006up}, BBO\,\cite{Yagi:2011wg}, ET\,\cite{Punturo_2010}, CE\,\cite{LIGOScientific:2016wof} and aLIGO \cite{LIGOScientific:2014pky} are shown as shaded regions. Finally, we show the parameter space in singlet VEV $u$, bias term $\Delta V$ and scale of leptogenesis $M_1$  in Fig.~\ref{fig:cntr-SNR} by keeping the CP phase parameter $\sin{2\phi} >0.1$. The points in these plots reflect the scale of leptogenesis shown in colour code as well as the signal-to-noise ratio (SNR) at respective GW experiments to be more than 10 where the SNR is defined as~\cite{Dunsky:2021tih, Schmitz:2020syl} 
\begin{equation}
\rho = \sqrt{\tau\,\int_{f_\text{min}}^{f_\text{max}}\,df\,\left[\frac{\Omega_\text{GW}(f)\,h^2}{\Omega_\text{expt}(f)\,h^2}\right]^2}\,, 
\end{equation}
with $\tau$ being the observation time for a particular detector. In each of these plots, the region above the coloured points are ruled out by BBN as well as CMB limits on $\Delta N_{\rm eff}$. On the other hand, the region below the coloured points correspond to SNR lower than 10. While we show the parameter space for GW experiments BBO \cite{Yagi:2011wg}, LISA\,\cite{AmaroSeoane2012LaserIS}, DECIGO \cite{Kawamura:2006up}, PPTA \cite{Manchester_2013}, IPTA \cite{Hobbs_2010},  EPTA \cite{Kramer_2013}, SKA \cite{Weltman:2018zrl}, THEIA \cite{Garcia-Bellido:2021zgu}, $\mu$ARES \cite{Sesana:2019vho} and NANOGrav \cite{McLaughlin:2013ira} only, for remaining experiments like ET, CE, GAIA the required SNR can not be obtained under the assumption that all the experiments will be operating for four years. Thus, a larger part of the parameter space remains within reach of low frequency GW experiments compared to the high frequency ones like LISA, ET, CE etc. 

\medskip
\noindent
\textbf{\textit{Conclusions:}} We proposed a novel way of probing Dirac leptogenesis via future observations of stochastic gravitational waves background generated by unstable domain walls in the early universe. Such walls arise due to spontaneous breaking of $Z_2$ symmetry which needs to be imposed in the minimal Dirac seesaw model to keep unwanted terms away from the interaction Lagrangian. A soft $Z_2$ breaking term creates a pressure difference across the domain walls. Such a pressure difference or bias term can make the walls unstable leading to the emission of GW and in the process the domain walls disappear without spoiling the success of standard cosmology. The GW amplitude depends crucially on this bias term, as well as the wall tension, that further depends on the scale of $Z_2$ symmetry breaking. The scale of $Z_2$ symmetry breaking, on the other hand, in the minimal setup, is the VEV of a scalar singlet, that dictates the scale of leptogenesis as it appears in the Type-I Dirac seesaw relation for light neutrino masses. This leads to interesting correlation between the scale of leptogenesis and DW tension (i.e., the singlet VEV) leading to potential GW detection prospects in several planned GW experimental facilities. We find, most of the future GW experiments (for example, the Pulsar Timing Array (PTA)), are likely to probe the parameter space of our framework corresponding to high scale Dirac leptogenesis, with improved sensitivity. While we kept the bias term independent in our analysis, considering explicit origin of such terms like from Planck suppressed operators $\Delta V \propto \eta^5/M_{\rm Pl}$~\cite{Rai:1992xw} can give stronger correlation between leptogenesis favoured parameter space and GW prospects. It may also be possible to have GW probe of intermediate scale leptogenesis, which corresponds to lower values of singlet VEV or DW tension, specially in GW experiments sensitive to smaller strain like SKA, THEIA, $\mu$ARES. However, such low or intermediate scale leptogenesis will involve a more detailed analysis including lepton flavour effects which is beyond the scope of present work. 
\medskip
\noindent
\acknowledgements
BB received funding from the Patrimonio Autónomo - Fondo Nacional de Financiamiento para la Ciencia, la Tecnología y la Innovación Francisco José de Caldas (MinCiencias - Colombia) grant 80740-465-2020. This project has received funding /support from the European Union's Horizon 2020 research and innovation programme under the Marie Sklodowska-Curie grant agreement No 860881-HIDDeN. 
\twocolumngrid

\end{document}